\documentstyle[amssymb,12pt,a4,sw20lart]{article}
\input tcilatex
\newcommand{\vol}{\mbox{$\cal V$}}
\begin{document}

\author{P.\ Gr\"{u}ter (1), F.\ Lalo\"{e} (1), A.E. Meyerovich (2) and W.\ Mullin
(3) \\
%EndAName
(1) Laboratoire Kastler Brossel de l'ENS\thanks{%
Laboratoire associ\'e au CNRS, UA 18, et \`a l'Universit\'e Pierre et Marie
Curie}\\
24 rue Lhomond, F 75005 Paris, France\\
(2) Department of Physics, University of Rhode Island\\
Kingston RI 02881, USA\\
(3) Hasbrouck Laboratory, University of Massachusetts\\
Amherst Mass 01003, USA}
\title{Ursell Operators in Statistical Physics III: thermodynamic properties of
degenerate gases }
\date{}
\maketitle

\begin{abstract}
We study in more detail the properties of the generalized Beth Uhlenbeck
formula obtained in a preceding article.\ This formula leads to a simple
integral expression of the grand potential of any dilute system, where the
interaction potential appears only through the matrix elements of the second
order Ursell operator $U_{2}$.\ Our results remain valid for significant
degree of degeneracy of the gas, but not when Bose Einstein (or BCS)
condensation is reached, or even too close to this transition point.\ We
apply them to the study of the thermodynamic properties of degenerate
quantum gases: equation of state, magnetic susceptibility, effects of
exchange between bound states and free particles, etc. We compare our
predictions to those obtained within other approaches, especially the
``pseudo potential'' approximation, where the real potential is replaced by
a potential with zero range (Dirac delta function).\ This comparison is
conveniently made in terms of a temperature dependent quantity, the ``Ursell
length'', which we define in the text.\ This length plays a role which is
analogous to the scattering length for pseudopotentials, but it is
temperature dependent and may include more physical effects than just binary
collision effects; for instance, for fermions at very low temperatures, it
may change sign or increase almost exponentially.\ As an illustration,
numerical results for quantum hard spheres are given.
\end{abstract}

\newpage \ 

\section{Introduction}

The use of quantum cluster expansions was introduced in 1938 by Kahn and
Uhlenbeck \cite{Kahn-1938}, who generalized to quantum statistical mechanics
the Ursell functions $U_q$ defined by this author in 1927 \cite{Ursell-1927}%
. The major virtue of cluster expansions is that they provide directly
density expansions for systems where the interaction potential is not a
necessarily small perturbation; in fact, it may even diverge at short
relative distances (hard cores for instance) while usual perturbations
theories generate power series in the interaction potential, where each term
becomes infinite for hard core potentials. Starting from a quantum cluster
analysis, the Beth Uhlenbeck formula \cite
{Beth-1937,Huang-1987-10.3,Pathria-1972-9.6} gives an explicit expression of
the first terms of a fugacity expansion (or virial expansion) for the grand
potential of a quantum gas.\ The expression is valid for any potential, the
latter being characterized by its phase shifts in a completely general way.

One should nevertheless keep in mind that the words ``density expansion''
have a double meaning in this context.\ In a dilute gas, there are actually
two dimensionless parameters which characterize ``diluteness'': the product $%
n^{1/3}b$, where $n$ is the number density of the gas and $b$ is a length
characterizing the potential range (diameter of hard cores for instance),
and the product $n^{1/3}\lambda _{T}$, where $\lambda _{T}$ is the quantum
thermal wavelength of the particles. The former parameter is small if,
classically, a snapshot of the system shows particles among which almost all
are moving freely, while the few that interact are engaged in binary
collisions only; the latter, purely quantum in nature, is sometimes called
the quantum degeneracy parameter, and remains small provided there is little
overlap of the quantum wave packets. The validity of the Beth Uhlenbeck
formula, as all fugacity expansions, therefore requires two independent
parameters to be small.

In a previous article \cite{I}, we discuss one method which conveniently
treats the two parameters separately, and allows one to include the effects
of statistics by exact summations while limiting the expansion to the lowest
orders in $n^{1/3}b$.\ The technique is based on the use of Ursell operators 
$U_{q}$ generalizing the Ursell functions (for a system of distinguishable
particles), coupled with the exact calculation of the effect of exchange
cycles $C_{l}$ of arbitrary length $l$; for short we call it the technique
of U-C diagrams. The result is another expression of the grand potential,
which is no longer a fugacity expansion since it includes a summation over
all sizes of exchange cycles so that statistical effects are included to all
orders.\ Truncating the expansion to its first terms (lowest $q$ values
and/or low order in a given $U_{q}$) gives results which remain valid for
``dilute degenerate systems'' \cite{FL}, that is for all systems where the
potential range is sufficiently small, but where the degeneracy of the
system may become significant\footnote{%
For bosons, Bose Einstein condensation is excluded since it requires a
summation over an infinite number of interaction terms, a question which we
will study in a forthcoming article}. In this article we will start from an
expression of the grand potential which is limited to the first order
correction in the second Ursell operator $U_{2}$, obtained in \cite{I} as a
trace over two particles of a product of operators. We reduce the trace to
an explicit integral where the effects of the interactions are contained in
a simple matrix element.\ The range of validity of our result is actually
similar to that of the calculations based on the use of pseudopotentials 
\cite{Huang-1987-10.5}, another approach where the final results
automatically include the summation of an infinite perturbation series in
terms of the initial potential.\ The two methods are comparable, but we
think that the U-C diagram method provides more general and more precise
results, basically because it includes the short range correlations between
the particles, and because all scattering channels with given angular
momentum as well as their exact energy dependence are included instead of
only one constant scattering length\footnote{%
This does not mean that one could not improve the theory of pseudopotentials
to include all phase shifts, since a general expression of the
pseudopotential is given by Huang in \cite{Huang-1957}, but to our knowledge
this has not been done explicitly.}. Another point of comparison is the
class of methods, for instance discussed in \cite{L-L-par6} or \cite{AGD},
where a renormalization procedure is used in order to obtain expansions in
terms of the scattering $T$ matrix instead of the interaction potential
itself $V$; in the calculations discussed in the present article, no
renormalization of this kind is needed since, roughly speaking, it is
already included in the $U_{q}$'s, which are the building blocks of our
method. Nevertheless, as we will see, our method is no longer valid when the
gaseous system is brought too close to a phase transition (superfluid
transition of single particles for bosons, of pairs for fermions).

We begin this article with a study of the expression of the grand potential,
and show how it can be expressed as an expression that is similar to the
well known Beth Uhlenbeck formula; actually it can be obtained from it by
two simple substitutions. We then discuss the physics contained in this
general result, as well as the changes introduced by the possible occurrence
of bound states.\ In particular we consider the effects of exchange between
bound and unbound particles, an effect which is not contained in the usual
Beth Uhlenbeck formalism; this kind of exchange may play some role in clouds
of laser cooled alkali atoms \cite{Cornell-1995,Bradley-1995,Davis-1995} for
which the potential is sufficiently attractive to sustain a large number of
bound states.\ In section \ref{spin1/2} we apply these results to spin 1/2
particles.\ Finally, in section \ref{elements}, we discuss the appropriate
quantities in terms of which one should describe the effects of the
potential on the physical properties of the system, and introduce for this
purpose the so called ``Ursell length'', which plays a role similar to the
scattering length $a$.\ In the theoretical study of quantum gases, and as
already mentioned, one frequently used method is to replace the real
interaction potential between the particles by a ``pseudopotential'' that
has no range (a Dirac delta function of the space variables), and to treat
this potential to first perturbation order; in other words one ignores the
distortion of the many-body wave functions at short relative distances and
the associated effects of the inter particle correlations.\ The
justification of this approach is based on the physical expectation that,
for a dilute gas, all the effects of the potential should be contained in
the binary collision phase shifts associated with the potential, which can
easily be reproduced to first order by a pseudopotential; meanwhile all
detailed information on the behavior of the wave function at short relative
distances can safely be discarded. Our formalism allows one to explicitly
distinguish between short range effects (``in potential effects'') and
asymptotic effects (out of the potential), which naturally leads to a
discussion of this ansatz.\ The interactions appear in terms of a matrix
element of an Ursell operator, which depends on the potential but does not
reduce to it; for fermions at low temperatures, the matrix elements contain
physical effects which are not included in usual treatments of normal Fermi
gases.

\section{The grand potential}

\subsection{Notation}

The basic object in terms of which most physical quantities will be written
in this article is the second Ursell operator $U_2$, defined by: 
\begin{equation}
U_2(1,2)=\left[ \text{e}^{-\beta H_2(1,2)}-\text{e}^{-\beta \left[
H_1(1)+H_1(2)\right] }\right]  \label{u2}
\end{equation}
where $H_1(1)$ and $H_1(2)$ are single particle hamiltonians, containing the
kinetic energy of the particle and, if necessary, its coupling to an
external potential, and where:

\begin{equation}
H_{2}(1,2)=H_{1}(1)+H_{1}(2)+V_{int}(1,2)  \label{h2}
\end{equation}
is the hamiltionan of two particles, including the mutual interaction
potential $V_{int}(1,2)$. Depending of the context, it may be more
convenient to use the symmetrized operator $U_{2}^{S,A}$: 
\begin{equation}
U_{2}^{S,A}(1,2)=U_{2}(1,2)\frac{\left[ 1+\eta P_{ex}\right] }{2}=\frac{%
\left[ 1+\eta P_{ex}\right] }{2}U_{2}(1,2)  \label{u2sa}
\end{equation}
where $P_{ex}$ is the exchange operator between particles 1 and 2 and $\eta $
has the value +1 for bosons, $-1$ for fermions. Moreover, the ``interaction
representation version'' of either $U_{2}$ and $U_{2}^{S,A}$, obtained by
multiplying these operators by $e^{\beta H_{1}(1)}e^{\beta H_{1}(2)}$, will
also be useful; we denote them with an additional bar over the operator, for
instance: 
\begin{equation}
\overline{U}_{2}(1,2)=\text{e}^{\beta H_{1}(1)}\text{e}^{\beta H_{1}(2)}%
\text{e}^{-\beta H_{2}(1,2)}-1  \label{u2bar}
\end{equation}
This operator and its symmetrized version $\overline{U}_{2}^{S,A}(1,2)$ act
only in the space of relative motion of the two particles; they have no
action at all on the variables of the center of mass.

Finally, if the particles are not submitted to an external potential ($H_1$
contains only the kinetic energy), it is convenient to introduce the
momentum ${\bf P}_G$ of the center of mass of the two particles as well as
the hamiltonian of the relative motion: 
\begin{equation}
H_{rel}=\frac{{\bf P}^2}m+V_{int}({\bf R})  \label{hrel}
\end{equation}
($m$ is the mass of the particles); then, $U_2(1,2)$ can be written in the
form of a product: 
\begin{equation}
U_2(1,2)=\text{e}^{-\beta \left( {\bf P}_G\right) ^2/4m}\times \left[
U_2(1,2)\right] _{rel}  \label{u2bis}
\end{equation}
with:

\begin{equation}
\left[ U_2(1,2)\right] _{rel}=\left[ \text{e}^{-\beta H_{rel}}-\text{e}%
^{-\beta \frac{{\bf P}^2}m}\right]  \label{u2rel}
\end{equation}

\subsection{Approximate expression of the thermodynamic potential}

We now start from relation (46) of \cite{I} which gives the grand potential
(multiplied by $-\beta $) in the form: 
\begin{equation}
\limfunc{Log}Z=\left[ \limfunc{Log}Z\right] _{ig}+\left[ \limfunc{Log}%
Z\right] _{int}  \label{1}
\end{equation}
where $\left[ \limfunc{Log}Z\right] _{ig}$ is the well known value of $%
\limfunc{Log}Z$ for the ideal gas: 
\begin{equation}
\begin{array}{cc}
\left[ \limfunc{Log}Z\right] _{ig} & =-\eta \text{Tr}\left\{ \limfunc{Log}%
\left[ 1-\eta z\text{e}^{-\beta H_{1}}\right] \right\} \\ 
& =\eta \text{Tr}\left\{ \limfunc{Log}\left[ 1+\eta f\right] \right\}
\end{array}
\label{1prime}
\end{equation}
and where the correction introduced by the interactions is: 
\begin{equation}
\left[ \limfunc{Log}Z\right] _{int}=z^{2}\,\text{Tr}_{1,2}\left\{
U_{2}^{S,A}(1,2)\,\left[ 1+\eta f(1)\right] \,\left[ 1+\eta f(2)\right]
\right\}  \label{1bis}
\end{equation}
In these equations, 
\begin{equation}
z=\text{e}^{\beta \mu }  \label{1ter}
\end{equation}
is the fugacity, $\beta =1/k_{B}T$ the inverse temperature and $\mu $ the
chemical potential, while $f$ is defined as the operator: 
\begin{equation}
f=\frac{z\text{e}^{-\beta H_{1}}}{1-\eta z\text{e}^{-\beta H_{1}}}  \label{3}
\end{equation}
Similar results can be found in the work of Lee and Yang, see formulas
(II.8) and (II.23) of ref. \cite{Lee-Yang2}. As mentioned in the
introduction, equation (\ref{1bis}) gives the correction introduced by the
interactions to the lowest order approximation in $U_{2}$; see \cite{I} for
a discussion of the higher order corrections. Using the definition of $%
\overline{U}_{2}^{S,A}$: 
\begin{equation}
\overline{U}_{2}^{S,A}(1,2)=\text{e}^{\beta H_{1}(1)}\text{e}^{\beta
H_{1}(2)}U_{2}^{S,A}(1,2)=\frac{1+\eta P_{ex}}{2}\left[ \text{e}^{\beta
H_{1}(1)}\text{e}^{\beta H_{1}(2)}\text{e}^{-\beta H_{2}(1,2)}-1\right]
\label{u2sabarre}
\end{equation}
as well as the relation $z$e$^{-\beta H_{1}}\left[ 1+\eta f\right] =f$, we
can rewrite (\ref{1bis}) in the form: 
\begin{equation}
\left[ \limfunc{Log}Z\right] _{int}=\,\text{Tr}_{1,2}\left\{ \overline{U}%
_{2}^{S,A}(1,2)\,f(1)f(2)\right\}  \label{logz-int}
\end{equation}
which expresses the correction as the average of the operator $\overline{U}%
_{2}^{S,A}(1,2)$ over unperturbed distributions functions $f$'s.

\subsection{Spinless particles and rotational invariance\label{rotation}}

For spinless particles, by making the trace in (\ref{logz-int}) explicit, we
obtain: 
\begin{equation}
\left[ \limfunc{Log}Z\right] _{int}=-\frac{\lambda _{T}^{2}%
%TCIMACRO{\TeXButton{V}{\vol} }
%BeginExpansion
\vol%
%EndExpansion
}{\left( 2\pi \right) ^{6}}\,\int d^{3}k_{1}\int d^{3}k_{2}\,f({\bf k}_{1})f(%
{\bf k}_{2})\,a_{U}^{S,A}(\left| {\bf k}_{1}-{\bf k}_{2}\right| )
\label{eq-base}
\end{equation}
where $%
%TCIMACRO{\TeXButton{V}{\vol}}
%BeginExpansion
\vol%
%EndExpansion
$ is the volume of the system, $\lambda _{T}$ the thermal wavelength: 
\begin{equation}
\lambda _{T}=\frac{h}{\sqrt{2\pi mk_{B}T}}  \label{lambda}
\end{equation}
and where $a_{U}^{S,A}(k)$ is defined by: 
\begin{equation}
a_{U}^{S,A}(k)=-\frac{%
%TCIMACRO{\TeXButton{V}{\vol} }
%BeginExpansion
\vol%
%EndExpansion
}{\lambda _{T}^{2}}<{\bf k}\mid \overline{U}_{2}^{S,A}\mid {\bf k}>=-\frac{%
%TCIMACRO{\TeXButton{V}{\vol} }
%BeginExpansion
\vol%
%EndExpansion
}{\lambda _{T}^{2}}e^{\beta \hbar ^{2}k^{2}/m}<{\bf k}\mid \left[
U_{2}^{S,A}\right] _{rel}\mid {\bf k}>  \label{sigmaprime}
\end{equation}
Here: 
\begin{equation}
{\bf k=}\frac{{\bf k}_{1}-{\bf k}_{2}}{2}  \label{krelatif}
\end{equation}
is the appropriate variable since $\overline{U}_{2}^{S,A}(1,2)$ does not
have any action in the space of states associated to the center of mass of
the two particles; $a_{U}^{S,A}(k)$ is a microscopic length, independent of $%
%TCIMACRO{\TeXButton{V}{\vol}}
%BeginExpansion
\vol%
%EndExpansion
$ for large systems\footnote{%
This is true since $\overline{U}_{2}^{S,A}(1,2)$ has a microscopic range and
since the factor $%
%TCIMACRO{\TeXButton{V}{\vol}}
%BeginExpansion
\vol%
%EndExpansion
$ in (\ref{sigmaprime}) makes up for the normalization factor of the plane
waves that occur in the matrix element (note that all plane wave kets in our
formulas are normalized in a finite volume; hence the absence of Dirac delta
functions of momenta differences in (\ref{sigmaprime}).}, which we will call
the ``Ursell length'' - see section \ref{Ur-length} for a more detailed
discussion and a justification of the numerical factors that we have
introduced. The correction written in (\ref{eq-base}) is analogous to a
first order energy correction due to binary interactions, while $%
a_{U}^{S,A}(k)$ plays the role of some effective interaction (within a
numerical factor).

For instance, if we assume that we treat to first order in perturbation a
pseudopotential\footnote{%
The most usual procedure is to treat this potential to first order only,
since a naive treatment of higher orders may introduce inconsistencies. For
instance, in three dimensions, it is possible to show that all phase shifts,
and therefore the collision cross section, of a zero range potential such as
(\ref{veff}), are exactly zero; on the other hand, they do not vanish to
first order (in other words, the Born series for potentials containing a
delta function is not convergent).\ For the same reason, for a potential
such as (\ref{veff}), the Ursell operator $U_{2}$ vanishes exactly, while it
does not if the potential is treated to first order.
\par
For a more elaborate discussion of pseudopotentials going beyond (\ref{veff}%
) and including waves of higher angular momentum, see \cite{Huang-1957}.} of
the form:

\begin{equation}
V_{eff}({\bf r})=\frac{4\pi \hbar ^{2}a}{m}\delta ({\bf r})  \label{veff}
\end{equation}
where $a$ is a scattering length (or the diameter of hard cores), we easily
obtain: 
\begin{equation}
<{\bf k}\mid \overline{U}_{2}^{S,A}\mid {\bf k}>\simeq -\frac{\lambda
_{T}^{2}}{%
%TCIMACRO{\TeXButton{V}{\vol} }
%BeginExpansion
\vol%
%EndExpansion
}a\left[ 1+\eta \right]  \label{elementsde}
\end{equation}
so that (\ref{sigmaprime}) shows that, in this approximation, the Ursell
length becomes independent of the wave number $k$.\ For bosons, the
correction to the grand potential then becomes: 
\begin{equation}
\left[ \limfunc{Log}Z\right] _{int}\simeq -2a\frac{%
%TCIMACRO{\TeXButton{V}{\vol} }
%BeginExpansion
\vol%
%EndExpansion
}{\lambda _{T}}\,\times g_{3/2}(z)  \label{variation}
\end{equation}
with the usual notation: 
\begin{equation}
g_{3/2}(z)=\frac{\lambda _{T}^{3}}{(2\pi )^{3}}\int d^{3}k\,\,f(k;z)
\label{defdeg}
\end{equation}
while, for fermions, no first order correction is obtained.\ These results
coincide with the first order terms of the well-known results of Lee and
Yang; see formulas (1) and (4) for $J=0$ of ref. \cite{LeeandYang}.\
Nevertheless, in section \ref{elements}, we discuss the validity of this
first order approximation in the calculation of the matrix elements of $%
U_{2} $ and conclude that, for bosons, (\ref{elementsde}) is a good
approximation while it is not necessarily the case for fermions.

At this point, it is convenient to introduce the free spherical waves $\mid
j_{k,l,m}^{(0)}>$ associated with the relative motion of the particles, with
wave functions\footnote{%
To normalize these functions (with a Dirac function of $k$ vectors and
Kronecker delta's of $l$ and $m$), it would be necessary to multiply all the 
$j_{k,l,m}^{(0)}({\bf r})$'s by factors $k\sqrt{2/\pi }$; this operation is
not necessary here.}: 
\begin{equation}
<{\bf r}\mid j_{k,l,m}^{(0)}>=j_{l}(kr)Y_{l}^{m}(\widehat{r})  \label{besel}
\end{equation}
(with standard notation; $j_{l}$ is a spherical Bessel function, $Y_{l}^{m}(%
\widehat{r})$ a spherical harmonics of the angular variables of ${\bf r}$);
if $\mid {\bf k>}$ is a plane wave normalized in a volume $%
%TCIMACRO{\TeXButton{V}{\vol}}
%BeginExpansion
\vol%
%EndExpansion
$: 
\begin{equation}
\mid {\bf k>=}\frac{4\pi }{\sqrt{%
%TCIMACRO{\TeXButton{V}{\vol} }
%BeginExpansion
\vol%
%EndExpansion
}}\sum_{l,m}(i)^{l}\left[ Y_{l}^{m}(\widehat{k})\right] ^{*}\mid
j_{k,l,m}^{(0)}>  \label{spherical}
\end{equation}
If we insert this equality into (\ref{sigmaprime}) and take into account the
well-known relation: 
\begin{equation}
\sum_{m}\left| Y_{l}^{m}(\widehat{k})\right| ^{2}=\frac{2l+1}{4\pi }
\label{normalis}
\end{equation}
we readily obtain the result\footnote{%
We assume rotational invariance, so that the matrix elements of $\overline{U}%
_2(1,2)$ are diagonal in $l$ and $m$.}: 
\begin{equation}
a_{U}^{S,A}(k)=\sum_{l}(2l+1)\left[ 1+\eta (-1)^{l}\right] \times
a_{U}^{(l)}(k)  \label{sigmaprim}
\end{equation}
with: 
\begin{equation}
a_{U}^{(l)}(k)=-\frac{2\pi }{\lambda _{T}^{2}}e^{\beta \hbar
^{2}k^{2}/m}<j_{klm}^{(0)}\mid \left[ U_{2}\right] _{rel}\mid j_{klm}^{(0)}>
\label{uldef}
\end{equation}
where $\left[ U_{2}(1,2)\right] _{rel}$ has been defined in (\ref{u2rel});
rotational invariance ensures that the right hand side of (\ref{uldef}) is
independent of $m$.\ These results can be inserted into the integral
appearing in (\ref{eq-base}) and provide an expression of the correction to
the grand potential which is a direct generalization of the usual Beth
Uhlenbeck formula to gases having a significant degree of degeneracy: 
\begin{equation}
\left[ \limfunc{Log}Z\right] _{int}=-\frac{\lambda _{T}^{2}%
%TCIMACRO{\TeXButton{V}{\vol} }
%BeginExpansion
\vol%
%EndExpansion
}{(2\pi )^{6}}\sum_{l}(2l+1)\left[ 1+\eta (-1)^{l}\right] \times \int
d^{3}k_{1}\int d^{3}k_{2}\,f({\bf k}_{1})f({\bf k}_{2})\times
a_{U}^{(l)}(k)\,  \label{bill}
\end{equation}

For comparison, we recall the explicit expression of this formula: 
\begin{equation}
\left[ \limfunc{Log}Z\right] _{int}^{B.U.}=\frac{2^{3/2}}{2\pi }z^{2}\frac{%
%TCIMACRO{\TeXButton{V}{\vol} }
%BeginExpansion
\vol%
%EndExpansion
}{\lambda _{T}}\sum_{l}(2l+1)\left[ 1+\eta (-1)^{l}\right] \int dk\text{e}%
^{-\beta \hbar ^{2}k^{2}/m}k\delta _{l}(k)  \label{bu}
\end{equation}
(we temporarily ignore possible bound states), where $\lambda _{T}$ is
defined in (\ref{lambda}). As already discussed in \cite{I}, if we replace
each of the two $f$'s in (\ref{bill}) by their low density limit $e^{\beta
(\mu -\hbar ^{2}k^{2}/2m)}$, the integration over $d^{3}K$ of a Gaussian
function introduces a factor $8^{3/2}\left[ \pi /\lambda _{T}\right] ^{3}$
and, after some algebra, we recover (\ref{bu}).\ In other words, the
following substitutions are necessary to obtain (\ref{bill}) from the usual
Beth Uhlenbeck formula \footnote{%
There are several equivalent ways to write the Beth Uhlenbeck formula; for
instance, an integration by parts allows one to replace the product $k\delta
_{l}(k)$ by the derivative $d\delta _{l}(k)/dk$ while the coefficient $%
1/\lambda _{T}$ is replaced by $\pi /\lambda _{T}^{3}$.\ Under these
conditions, the second line of (\ref{subst})becomes $d\delta
_{l}(k)/dk\Rightarrow -(k\lambda _{T})^{2}a_{u}^{(l)}(k)/\pi $. In other
words, the correspondence between ou result and the Beth Uhlenbeck formula
depends on the way the latter is written.
\par
The two functions $k\delta _{l}(k)$ and $-k^{2}a_{U}^{(l)}(k)$ are not
necessarily equal but, when multiplied by a Gaussian function $e^{-\beta
\hbar ^{2}{\bf k}^{2}/m}$, have the same integral over $d^{3}k$.}: 
\begin{equation}
\left\{ 
\begin{array}{c}
\displaystyle \text{e}^{-\beta \hbar \left[ \left( {\bf k}_{1}\right)
^{2}+\left( {\bf k}_{1}\right) ^{2}\right] /2m}\Rightarrow z^{-2}f({\bf k}%
_{1})f({\bf k}_{2}) \\ 
\displaystyle \delta _{l}(k)\Rightarrow -ka_{U}^{(l)}(k)
\end{array}
\right.  \label{subst}
\end{equation}
In equation (\ref{1bis}), the first of these substitutions amounts to adding
the terms in $\eta f$.

\subsection{Bound states}

The operator $\left[ U_{2}\right] _{rel}^{S,A}$ may be written as the sum of
the contributions of bound states and of the continuum: 
\begin{equation}
\left[ U_{2}\right] _{rel}^{S,A}=\sum_{n}\mid \Phi _{n}\rangle \langle \Phi
_{n}\mid \text{e}^{-\beta E_{n}}+\text{continuum}  \label{decomposition}
\end{equation}
where the $\mid \Phi _{n}\rangle $'s are the kets associated with the
eigenstate of the relative motion of the particles with (negative) energy $%
-E_{n}$ (with appropriate symmetry for the statistics of the particles).\ It
is therefore not difficult make the contribution of bound states in (\ref
{eq-base}) explicit, which provides the following term: 
\begin{equation}
\begin{array}{ll}
\displaystyle\left[ \limfunc{Log}Z\right] _{int}^{bound} & =\displaystyle%
(2\pi )^{-6}%
%TCIMACRO{\TeXButton{V}{\vol} }
%BeginExpansion
\vol%
%EndExpansion
^{2}\int d^{3}K\int d^{3}q\,\,f(\frac{{\bf K}}{2}+{\bf q)}\,f(\frac{{\bf K}}{%
2}-{\bf q)\times } \\ 
& \displaystyle\,\,\,\,\,\,\,\,\;\;\;\;\;\;\times \,\,\QATOPD. . {\QATOPD. .
{} {}} {}\sum_{n}^{\QATOPD. . {\QATOPD. . {} {}} {}}\mid \langle {\bf q}\mid
\Phi _{n}\rangle \mid ^{2}\text{e}^{\beta \left( E_{n}+\hbar
^{2}q^{2}/m\right) }
\end{array}
\label{int-bound}
\end{equation}
where ; because these kets have a finite range, and because the plane waves $%
\mid {\bf q}>$ are normalized in a macroscopic volume $%
%TCIMACRO{\TeXButton{V}{\vol}}
%BeginExpansion
\vol%
%EndExpansion
$, the product $%
%TCIMACRO{\TeXButton{V}{\vol}}
%BeginExpansion
\vol%
%EndExpansion
\mid \langle {\bf q}\mid \Phi _{n}\rangle \mid ^{2}$ is independent of $%
%TCIMACRO{\TeXButton{V}{\vol}}
%BeginExpansion
\vol%
%EndExpansion
$ in the thermodynamic limit, as necessary to obtain an extensive correction
to the grand potential.\ If we rewrite the integral of (\ref{int-bound}) in
the form: 
\begin{equation}
\int d^{3}K\int d^{3}q\,e^{-\beta \hbar ^{2}{\bf K}^{2}/4m}\left[ 1+\eta f(%
\frac{{\bf K}}{2}+{\bf q)}\right] \times \left[ 1+\eta f(\frac{{\bf K}}{2}-%
{\bf q)}\right] \sum_{n}\mid \langle {\bf q}\mid \Phi _{n}\rangle \mid ^{2}%
\text{e}^{\beta E_{n}}  \label{integr}
\end{equation}
we may distinguish between two contributions in the correction: 
\begin{equation}
\left[ \limfunc{Log}Z\right] _{int}^{bound}=\left[ \limfunc{Log}Z\right]
_{B.U.}^{bound}+\left[ \limfunc{Log}Z\right] _{stat.}^{bound}  \label{somme}
\end{equation}
The first contribution is obtained by ignoring in (\ref{integr}) the $\eta f$%
's inside the brackets, which allows one to integrate over $d^{3}K$; using
the closure relations over the plane waves $\mid {\bf q}\rangle $ and the
normalization of the bound states $\mid \Phi _{n}\rangle $ then provides the
following result for this first contribution of the bound states: 
\begin{equation}
\left[ \limfunc{Log}Z\right] _{B.U.}^{bound}=2^{3/2}\left[ \lambda
_{T}\right] ^{-3}%
%TCIMACRO{\TeXButton{V}{\vol} }
%BeginExpansion
\vol%
%EndExpansion
\sum_{n}\,\text{e}^{\beta E_{n}}  \label{bound-bu}
\end{equation}
which is identical to the term corresponding to bound states in the usual
Beth Uhlenbeck formula. The second contribution arises from the rest of the
product of the brackets and is equal to: 
\begin{equation}
\begin{array}{cl}
\displaystyle\left[ \limfunc{Log}Z\right] _{stat.}^{bound} & \displaystyle%
=(2\pi )^{-6}%
%TCIMACRO{\TeXButton{V}{\vol} }
%BeginExpansion
\vol%
%EndExpansion
^{2}\int d^{3}K\int d^{3}q\,\text{e}^{-\beta \hbar ^{2}{\bf K}^{2}/4m}\left[
f(\frac{{\bf K}}{2}+{\bf q)}f(\frac{{\bf K}}{2}-{\bf q)}+\right. \\ 
& \displaystyle\;\;\;\left. +\eta f(\frac{{\bf K}}{2}+{\bf q)+}\eta f(\frac{%
{\bf K}}{2}-{\bf q)}\right] \sum_{n}\mid \langle {\bf q}\mid \Phi
_{n}\rangle \mid ^{2}\text{e}^{\beta E_{n}}
\end{array}
\label{bound-stat}
\end{equation}
It accounts for exchange effects between bound states and continuum states%
\footnote{%
The exchange effects between bound states themselves will be investigated in
another article with the study of pair condensation (BCS condensation for
fermions); they are higher order in $U_2$.}, which explains the appearance
of the scalar product $\langle {\bf q}\mid \Phi _{n}\rangle $ of a free
plane wave and a bound state wave function; the factors $e^{\beta E_{n}}$
correspond physically to the Boltzmann distribution of the populations of
the bound states.

If we assume that the momenta in the continuum, which have values that are
limited by the presence of the $f$'s under the integral, are much smaller
than those contained in the bound states $\mid \Phi _{n}\rangle $'s, we can
replace the product $\langle {\bf q}\mid \Phi _{n}\rangle $ by $\langle {\bf %
q=0}\mid \Phi _{n}\rangle $; this shows that the effect of statistics is
more important for bound states with a wave function with a significant
integral over space, as for instance the ground state wave function which
has no node; states with rapidly oscillating wave functions give almost no
correction to the usual Beth Uhlenbeck formula. Generally speaking, for
bosons, the effect of statistics is always to increase the Beth Uhlenbeck
term (\ref{bound-bu}). For fermions, the situation is more complicated: if
the gas is only slightly degenerate one has $f^{2}<f$ so that the terms in $%
\eta $ dominate in (\ref{bound-stat}), leading to a decrease of the effects
of bound states; but if the gas is strongly degenerate, there seems to be no
general rule, and exchange of bound states with the continuum may either
enhance or reduce their contribution.

\subsection{Spins\label{spins}}

An easy generalization is to include spins; the only difference is that all
the traces must now also include spin states.\ It is then convenient to
replace the functions $f({\bf k})$ of momentum by spin operators $f_{S}({\bf %
k})$ (corresponding to two by two matrices for spin 1/2 particles) which are
functions of ${\bf k}$ and are defined as: 
\begin{equation}
\langle m_{S}\mid f_{S}({\bf k})\mid m_{S}^{^{\prime }}\rangle =\langle
m_{S},{\bf k}\mid \frac{ze^{-\beta H_{1}}}{1-\eta ze^{-\beta H_{1}}}\mid
m_{S}^{^{\prime }},{\bf k\rangle }  \label{rho}
\end{equation}
where $H_{1}$ is the one particle hamiltonian, including kinetic energy as
well as coupling of the spins to the magnetic field (if the particles carry
magnetic moments).\ Similarly, because now the exchange of particles must
also include their spin states, $a_{U}^{S,A}(k)$ becomes an operator $\Sigma
_{S}(k)$ which acts in the space of the states of two spins: 
\begin{equation}
\Sigma _{S}(k)=\sum_{l=0}^{\infty }(2l+1)\,\,a_{U}^{(l)}(k)\left[ 1+\eta
(-1)^{l}\,P_{ex}^{S}\right]  \label{sigma-spins}
\end{equation}
where $P_{ex}^{S}$ is the exchange operators of two spins.\ This leads to
the following generalization of (\ref{eq-base}): 
\begin{equation}
\left[ \limfunc{Log}Z\right] _{int}=-\frac{\lambda _{T}^{2}%
%TCIMACRO{\TeXButton{V}{\vol} }
%BeginExpansion
\vol%
%EndExpansion
}{\left( 2\pi \right) ^{6}}\int d^{3}k_{1}\int d^{3}k_{2}\;\text{Tr}%
_{S_{1},S_{2}}\left\{ f_{S}({\bf k}_{1})f_{S}({\bf k}_{2})\;\,\Sigma
_{S}(k)\right\}  \label{Z-spins}
\end{equation}

\section{Spin 1/2 particles\label{spin1/2}}

We now apply the preceding calculation to the study of the magnetic
susceptibility of a dilute gas of fermions or bosons with spin 1/2 (spin
polarized atomic hydrogen provides an example of spin 1/2 bosons \cite
{Greytak-1984, Silvera-1986}). We assume that the one particle hamiltonian
is: 
\begin{equation}
H_{1}=\frac{{\bf P}^{2}}{2m}-\frac{\hbar \omega _{0}}{2}\sigma _{z}
\label{h1}
\end{equation}
where $\omega _{0}/2\pi $ is the Larmor frequency in the (homogeneous)
magnetic field and $\sigma _{z}$ the Pauli matrix (operator) associated with
the component of the spin along the field; we set: 
\begin{equation}
z_{\pm }=\text{e}^{\beta \left( \mu \pm \hbar \omega _{0}/2\right) }=z\text{e%
}^{\pm \beta \hbar \omega _{0}/2}  \label{zplus}
\end{equation}
which gives the following values for the matrix elements\footnote{%
We remind the reader that, except for an ideal gas, the $f_{\pm }$'s are not
the populations of the one body density operators, but differ from them by
density corrections \cite{II}.} of $f_{S}({\bf k})$ defined in (\ref{rho}): 
\begin{equation}
f_{\pm }({\bf k})=\frac{z_{\pm }\text{e}^{-\beta \hbar ^{2}k^{2}/2m}}{1-\eta
z_{\pm }\text{e}^{-\beta \hbar ^{2}k^{2}/2m}}  \label{fplus}
\end{equation}
For short we will write: 
\begin{equation}
\begin{array}{cc}
f_{\pm }(1)=f_{\pm }({\bf k}_{1})\,\,\,\,\,\,\,\,\,\,\,\,\, & 
\,\,\,\,\,\,\,\,\,\,\,\,\,\,\,\,\,\,\,f_{\pm }(2)=f_{\pm }({\bf k}_{2})
\end{array}
\label{f12}
\end{equation}
With this notation we have: 
\begin{equation}
\begin{array}{cc}
\displaystyle\left[ \limfunc{Log}Z\right] _{ig} & \displaystyle=-\eta \frac{%
%TCIMACRO{\TeXButton{V}{\vol} }
%BeginExpansion
\vol%
%EndExpansion
}{8\pi ^{3}}\int d^{3}k_{1}\limfunc{Log}\left\{ \stackrel{}{\left[ 1-\eta
z_{+}\text{e}^{-\beta \hbar ^{2}k_{1}^{2}/2m}\right] }\left[ 1-\eta z_{-}%
\text{e}^{-\beta \hbar ^{2}k_{1}^{2}/2m}\right] \right\} \\ 
& \displaystyle=\eta \frac{%
%TCIMACRO{\TeXButton{V}{\vol} }
%BeginExpansion
\vol%
%EndExpansion
}{8\pi ^{3}}\int d^{3}k_{1}\limfunc{Log}\left\{ \stackrel{}{\left[ 1+\eta
f_{+}(1)\right] }\left[ 1+\eta f_{-}(1)\right] \right\}
\end{array}
\label{logzig}
\end{equation}
For a gas at equilibrium the operators $f_{S}({\bf k})$ are diagonal in the
basis corresponding to a quantization axis parallel to the magnetic field;
we then have: 
\begin{equation}
\begin{array}{cl}
\displaystyle\text{Tr}_{S_{1},S_{2}}\left\{ \stackrel{}{f_{S}({\bf k}%
_{1})f_{S}({\bf k}_{2})}\right\} & \displaystyle=\text{Tr}_{S_{1}}\left\{
f_{S}({\bf k}_{1})\right\} \times \text{Tr}_{S_{2}}\left\{ f_{S}({\bf k}%
_{2})\right\} \\ 
& \displaystyle=\,\stackrel{\stackrel{}{}}{\left[ f_{+}(1)+f_{-}(1)\right] }%
\left[ f_{+}(2)+f_{-}(2)\right]
\end{array}
\label{trs1s2}
\end{equation}
as well as: 
\begin{equation}
\text{Tr}_{S_{1},S_{2}}\left\{ f_{S}({\bf k}_{1})f_{S}({\bf k}%
_{2})P_{ex}^{S}\right\} =f_{+}(1)f_{+}(2)+f_{-}(1)f_{-}(2)  \label{trex}
\end{equation}
(the latter result arises because the trace gets non zero contributions only
from the two spin states $\mid +,+\rangle $ and $\mid -,-\rangle $, which
are invariant under the effect of $P_{ex}^{S}$).\ We therefore have: 
\begin{equation}
\begin{array}{cc}
\displaystyle\left[ \limfunc{Log}Z\right] _{int} & \displaystyle=-\frac{%
\lambda _{T}^{2}%
%TCIMACRO{\TeXButton{V}{\vol} }
%BeginExpansion
\vol%
%EndExpansion
}{\left( 2\pi \right) ^{6}}\int d^{3}k_{1}\int d^{3}k_{2}\,\left\{
a_{U}^{S,A}(k)\,\stackrel{\stackrel{}{}}{\left[
f_{+}(1)f_{+}(2)+f_{-}(1)f_{-}(2)\right] }\right. \\ 
& \displaystyle\,\,\,\,\,\,\,\,\,\,\,\,\,\,\,\,\,\,\,\,\,\,\,\,\,\,\left.
+2a_{U}(k)\;f_{+}(1)f_{-}\stackrel{\stackrel{}{}}{(2)}\right\}
\end{array}
\label{logzint}
\end{equation}
where: 
\begin{equation}
k=\frac{\mid {\bf k}_{1}-{\bf k}_{2}\mid }{2}  \label{k}
\end{equation}
while $a_{U}^{S,A}(k)$ is defined by (\ref{sigmaprime}) - or equivalently (%
\ref{sigmaprim}) - while $a_{U}(k)$ is the un-symmetrized version\footnote{%
Note the factor 1/2 which does not appear in (\ref{sigmaprime}); we choose
this convention since the same factor 1/2 appears in the definition (\ref
{u2sa}) of the symmetrized version of $U_{2}$; in this way, if exchange
effects are ignored (high temperature limit for instance), the various
Ursell lengths become equal.} of the Ursell length: 
\begin{equation}
a_{U}(k)=-\frac{%
%TCIMACRO{\TeXButton{V}{\vol} }
%BeginExpansion
\vol%
%EndExpansion
}{2\lambda _{T}^{2}}<{\bf k}\mid \overline{U}_{2}\mid {\bf k}%
>=\sum_{l=0}^{\infty }(2l+1)a_{U}^{(l)}(k)\,\,  \label{sigma}
\end{equation}
(it would correspond to distinguishable particles). If, as in the beginning
of section \ref{rotation}, we treat the pseudopotential (\ref{veff}) to
first order, for fermions we obtain $a_{U}^{A}(k)=0$ and $a_{U}(k)=a$; the
first order correction is now given by: 
\begin{equation}
\left[ \limfunc{Log}Z\right] _{int}\simeq -2\frac{a%
%TCIMACRO{\TeXButton{V}{\vol} }
%BeginExpansion
\vol%
%EndExpansion
}{\lambda _{T}}\,\times g_{3/2}(-z)  \label{corrferm}
\end{equation}
which coincides with the result of Lee and Yang (equation (1) of \cite
{LeeandYang} for $J=1/2$); but, again, we note that a critical discussion of
this first order calculation is made in section \ref{elements} which shows
that, for fermions, the results of the Ursell approach may be different from
those of a pseudopotential theory.

\subsection{Density of particles and of energy; magnetization.}

The number density of the gas is obtained from the relation: 
\begin{equation}
n=\frac{\langle N\rangle }{%
%TCIMACRO{\TeXButton{V}{\vol} }
%BeginExpansion
\vol%
%EndExpansion
}=%
%TCIMACRO{\TeXButton{V}{\vol} }
%BeginExpansion
\vol%
%EndExpansion
^{-1}\;z\frac{\partial }{\partial z}[\limfunc{Log}Z]  \label{n}
\end{equation}
Similarly, then density of internal energy is given by: 
\begin{equation}
w=\frac{\langle U\rangle }{%
%TCIMACRO{\TeXButton{V}{\vol} }
%BeginExpansion
\vol%
%EndExpansion
}=-%
%TCIMACRO{\TeXButton{V}{\vol} }
%BeginExpansion
\vol%
%EndExpansion
^{-1}\frac{\partial }{\partial \beta }[\limfunc{Log}Z]  \label{u}
\end{equation}
while the ``magnetization''\footnote{%
What we call here magnetization is not a real magnetic moment (ampere square
meter) but a dimensionless number equal to the sum of the average values of $%
\sigma _z$ of all atoms; in other words, the maximum value of $M$ (complete
spin polarization, all spins parallel) is equal to the total number of
particles.} is equal to: 
\begin{equation}
M=\frac{2}{\beta \hbar }\frac{\partial }{\partial \omega _{0}}[\limfunc{Log}Z%
]  \label{m}
\end{equation}
We therefore have to vary either $z$, or $\beta $, or $\omega _{0}$ in
formulas (\ref{logzig}) and (\ref{logzint}); we can then use the simple
relations: 
\begin{equation}
dz_{\pm }=z_{\pm }\left( \frac{dz}{z}\pm \frac{\hbar \omega _{0}}{2}d\beta
\pm \beta \hbar \frac{d\omega _{0}}{2}\right)  \label{dz}
\end{equation}
and: 
\begin{equation}
z_{\pm }\frac{\partial f_{\pm }}{\partial z_{\pm }}=f_{\pm }\left[ 1+\eta
f_{\pm }\right]  \label{derf}
\end{equation}
as well as: 
\begin{equation}
\frac{\partial f_{\pm }}{\partial \beta }=f_{\pm }\left[ 1+\eta f_{\pm
}\right] \left[ -\frac{\hbar ^{2}k^{2}}{2m}\pm \frac{\hbar \omega _{0}}{2}%
\right]  \label{dfdbeta}
\end{equation}
We then obtain, by making use of the symmetry in the indices $1$ and $2$: 
\begin{equation}
\begin{array}{cl}
\displaystyle n & \displaystyle=(2\pi )^{-3}\int d^{3}k_{1}\left[
f_{+}(1)+f_{-}(1)\right] + \\ 
& \displaystyle-\frac{\lambda _{T}^{2}}{\left( 2\pi \right) ^{6}}\int
d^{3}k_{1}\int d^{3}k_{2}\,\left\{ 2a_{U}^{S,A}(k)f_{+}(1)f_{+}(2)\stackrel{%
\stackrel{}{}}{\left[ 1+\eta f_{+}(1)\right] }+id.\left(
f_{+}\Leftrightarrow f_{-}\right) +\right. \\ 
& \,\,\,\,\,\,\,\,\,\,\,\,\,\,\,\,\,\,\,\,\,\,\,\,\,\,\,\,\,\,\,\,\,\,\,%
\displaystyle\,\,\,\,\,\left. +2a_{U}(k)f_{+}(1)f_{-}(2)\stackrel{\stackrel{%
}{}}{\left[ 2+\eta f_{+}(1)+\eta f_{-}(2)\right] }\right\}
\end{array}
\label{n2}
\end{equation}
where $id.\left( f_{+}\Leftrightarrow f_{-}\right) $ symbolizes the same
expression where $f_{+}$ and $f_{-}$ are interchanged. The calculation of $M 
$ is almost the same, except that now, because of (\ref{dz}), $f_{+}$ and $%
f_{-}$ introduce different signs; the result is: 
\begin{equation}
\begin{array}{cl}
\displaystyle \frac{M}{%
%TCIMACRO{\TeXButton{V}{\vol} }
%BeginExpansion
\vol%
%EndExpansion
} & \displaystyle=(2\pi )^{-3}\int d^{3}k_{1}\left[ f_{+}(1)-f_{-}(1)\right]
+ \\ 
& \displaystyle-\frac{\lambda _{T}^{2}}{\left( 2\pi \right) ^{6}}\int
d^{3}k_{1}\int d^{3}k_{2}\,\left\{ 2a_{U}^{S,A}(k)f_{+}(1)f_{+}(2)\stackrel{%
\stackrel{}{}}{\left[ 1+\eta f_{+}(1)\right] }-id.\left(
f_{+}\Leftrightarrow f_{-}\right) +\right. \\ 
& \,\,\,\,\,\,\,\,\,\,\,\,\,\,\,\,\,\,\,\,\,\,\,\,\,\,\,\,\,\,\,\,\,\,\,%
\displaystyle\,\,\,\,\,\left. +2\eta a_{U}(k)f_{+}(1)f_{-}(2)\stackrel{%
\stackrel{}{}}{\left[ f_{+}(1)-f_{-}(2)\right] }\right\}
\end{array}
\label{M}
\end{equation}
Finally, the calculation of the internal energy $w$ provides the result: 
\begin{equation}
w=w_{E}+w_{M}  \label{usomme}
\end{equation}
where $w_{E}$ is the density of energy associated with the external
variables of the particles (kinetic and potential energy):

\begin{equation}
\begin{array}{cl}
\displaystyle w_{E} & \displaystyle=(2\pi )^{-3}\int d^{3}k_{1}\left[
f_{+}(1)+f_{-}(1)\right] \frac{\hbar ^{2}k_{1}^{2}}{2m}+ \\ 
& \displaystyle-\frac{\lambda _{T}^{2}}{\left( 2\pi \right) ^{6}}\int
d^{3}k_{1}\int d^{3}k_{2}\,\left\{ 2a_{U}^{S,A}(k)f_{+}(1)f_{+}(2)\left[
1+\eta f_{+}(1)\right] \frac{\hbar ^{2}k_{1}^{2}}{2m}+id.\left(
f_{+}\Leftrightarrow f_{-}\right) +\right. \\ 
& \,\,\,\,\,\,\,\,\,\,\,\,\,\,\,\,\,\,\,\,\,\,\,\,\,\,\,\,\,\,\,\,\,\,\,%
\displaystyle\,\,\,\,\,\left. +2a_{U}(k)f_{+}(1)f_{-}(2)\left[ \left[ 1+\eta
f_{+}(1)\right] \frac{\hbar ^{2}k_{1}^{2}}{2m}+\left[ 1+\eta f_{-}(2)\right] 
\frac{\hbar ^{2}k_{2}^{2}}{2m}\right] \right\}
\end{array}
\label{uk}
\end{equation}
while $w_{M}$ is the density of magnetic energy: 
\begin{equation}
w_{M}=-\frac{\hbar \omega _{0}}{2}\frac{M}{%
%TCIMACRO{\TeXButton{V}{\vol} }
%BeginExpansion
\vol%
%EndExpansion
}  \label{um}
\end{equation}

We now assume that the magnetic field is sufficiently low so that the
preceding expressions may be limited to their first order expansion in $%
\omega _{0}\simeq 0$.\ We then use the relations: 
\begin{equation}
f_{\pm }\simeq f+\frac{\partial f_{\pm }}{\partial z_{\pm }}\,\frac{\partial
z_{\pm }}{\partial \omega _{0}}\omega _{0}=f\pm z\frac{\partial f}{\partial z%
}\,\frac{\beta \hbar \omega _{0}}{2}=f\left[ 1\pm \beta \hbar \frac{\omega
_{0}}{2}\left[ 1+\eta f\right] \right]  \label{df}
\end{equation}
which provide: 
\begin{equation}
\begin{array}{cl}
\displaystyle n(\omega _{0})\simeq n(0) & \displaystyle=(4\pi ^{3})^{-1}\int
d^{3}k_{1}f(1)-\frac{\lambda _{T}^{2}}{\left( 2\pi \right) ^{6}}\int
d^{3}k_{1}\int d^{3}k_{2}f(1)f(2)\times \\ 
& \,\displaystyle\,\,\,\,\,\,\,\,\,\,\,\,\,\,\,\,\times \stackrel{}{4\left[
1+\eta f(1)\right] }\left[ a_{U}^{S,A}(k)+a_{U}(k)\right]
\end{array}
\label{nchampnul}
\end{equation}
and for the zero field magnetic susceptibility of the gas: 
\begin{equation}
\begin{array}{cl}
\displaystyle\chi =\frac{M}{d\omega _{0}%
%TCIMACRO{\TeXButton{V}{\vol} }
%BeginExpansion
\vol%
%EndExpansion
} & \displaystyle=\beta \hbar \int d^{3}k_{1}f(1)\left[ 1+\eta f(1)\right] -%
\frac{\lambda _{T}^{2}}{\left( 2\pi \right) ^{6}}\beta \hbar \int
d^{3}k_{1}\int d^{3}k_{2}f(1)f(2)\times \\ 
& \displaystyle\times \left\{ 2\,a_{U}^{S,A}(k)\,\stackrel{\stackrel{}{}}{%
\left[ 1+\eta f(1)\right] }\left[ 2+2\eta f(1)+\eta f(2)\right] \right. + \\ 
& \,\,\,\,\,\,\displaystyle\,\,\,\,\,\,\,\,\,\,\,\,\,\,\,\,\,\,\,\,\,\,\,\,%
\,\,\,\,\,\,\,\,\,\,\,+\left. a_{U}(k)\stackrel{}{\left[ \left[
f(1)-f(2)\right] ^{2}+\,\,2\eta f(1)\left[ 1+\eta f(1)\right] \right] }%
\right\}
\end{array}
\label{chi}
\end{equation}

\subsection{Physical discussion\label{physdisc}}

Equations (\ref{n2}) and (\ref{M}) show how quantum statistical effects
modify the equation of state and the magnetic susceptibility of a gas. What
determines the properties of the dilute system is not directly the matrix
elements of the potential $V_{int}$, but those of the operator $\overline{U}%
_{2}$. Of course, if the potential is weak, the calculation may be limited
to first order perturbation theory, so that $\overline{U}_{2}$ may be
replaced by $-\beta V_{int}$; the two operators then become equivalent.\
But, in more realistic situations, for instance in atomic systems where the
interaction potential at short relative distances becomes very large, the
effects of the potential cannot be treated to first order.\ An obvious
difference between $\overline{U}_{2}$ and $V_{int}$ is that the former
depends on one more characteristic length than the potential, namely the
thermal wavelength $\lambda _{T}$, so that one can expect that its range
will depend explicitly on the temperature. It is sometimes argued that a
description of the interactions in terms of the scattering length is
sufficient at very low temperatures, because the range of the potential $%
V_{int}$ remains always much smaller than the de Broglie wavelengths of the
particles.\ As far as the matrix elements of $\overline{U}_{2}$ are
concerned, this is of course true if the interaction potential is treated to
first order only but, precisely at very low temperatures, the higher order
terms become significant and $\overline{U}_{2}$ may acquire a range that
exceeds by far that of the potential itself.\ We will see examples of this
phenomenon in section \ref{elements}.

We now discuss the limit of validity of the results obtained in the
preceding section or, since they are equivalent, of equations (\ref{logz-int}%
) and (\ref{Z-spins}).\ If the particles are bosons, our theory is limited
to gases which are not too close to Bose Einstein condensation - while of
course the degeneracy may be much more pronounced than for the Beth
Uhlenbeck formula since $z$ does not have to be small.\ The reason is that,
when a system of bosons approaches the region of quantum condensation, the
distribution function $f$ starts to build up a singularity at low energies.
One can then see that, when $z\rightarrow 1$, the partition function becomes
more and more sensitive to terms of higher and higher order in $\overline{U}%
_{2}$, in $\overline{U}_{3}$, $\overline{U}_{4\text{, }}$etc.\ , since they
contain larger and larger number of functions $f$ (or of factors $1+\eta f)$%
.\ In other words, terms which normally remain small corrections become
dominant when the point of Bose Einstein condensation is approached; see the
discussion already given in \cite{FL}, where it is emphasized\footnote{%
Unfortunately, in \cite{FL} an assumption of this discussion is not made
explicit, namely the fact that the matrix elements of $U_2$ should be
positive (dominant character of the attractive interactions in the matrix
element).} that a theory limited to first order in $\overline{U}_{2}$ would
predict the disappearance of the Bose Einstein condensation phenomenon and
its replacement by a simple crossover between two regimes.

For fermions, the discussion is different since no special phenomenon takes
place when $z$ reaches one; when the gas is cooled at constant density, this
merely corresponds to a cross over region where the gas is becoming
degenerate. When the temperature is decreased even more, a stronger
degeneracy builds up while $z=$e$^{\beta \mu }$ becomes larger and larger ($%
\beta $ increases while $\mu $ remains almost constant if the density is
fixed).\ It is therefore clear that $z$ itself can not be an expansion
parameter in this region; but our perturbation series is not a $z$ expansion
and it may still converge for degenerate systems, provided the matrix
elements of $\overline{U}_{2}$ are sufficiently small. In the next section,
we will see that these matrix elements are equal to the product $\lambda
_{T}^{2}a_{U}$, where $\lambda _{T}$ is the thermal wave length and $a_{U}$
is some microscopic length (the Ursell length) that we will define more
precisely later.\ Let us for instance discuss (\ref{n2}) where the
interaction corrections involve a double integral over two momenta.\ The
first introduces the number density $n$; the second contains a function $%
f(1-f)$, which in the limit of low temperatures is non zero only in an
energy slice of width $\beta ^{-1}$ at the surface of the Fermi sphere; the
corresponding width $\Delta k$ in terms of momentum is given by: 
\begin{equation}
\frac{\hbar ^{2}k_{F}\Delta k}{m}=\beta ^{-1}  \label{deltak}
\end{equation}
(where $\hbar k_{F}$ is the momentum at the Fermi surface) so that the
result of these integrations is: 
\begin{equation}
n\left( \lambda _{T}^{2}a_{U}\right) \times 4\pi k_{F}^{2}\frac{2\pi }{%
k_{F}\lambda _{T}^{2}}=2nk_{F}a_{U}  \label{2pi}
\end{equation}
The small parameter of the expansion\footnote{%
We note in passing that the phase occupation factor $(1-f)$ plays an
essential role in this argument.\ This factor occurs in all expressions, for
instance, in (\ref{M}), the bracket $\left[ f_{+}(1)-f_{-}(2)\right] $ may
be written as $\left[ f_{+}(1)-1+1-f_{-}(2)\right] $. If we had ignored this
factor, we would have found $na_{U}\lambda _{T}^{2}$ as the expansion
parameter, which is a temperature dependent factor (as in the usual Beth
Uhlenbeck formula).} is therefore the product $k_{F}a_{U}$, which is not
very different from $n^{1/3}a_{U}$ for degenerate gases, and indeed remains
small if the average distance between the particles is much larger than $%
a_{U}$. This result is not surprising: indeed, in the usual theory of dilute
non ideal Fermi gases \cite{L-L-par6}, the expansion parameter is
independent of the temperature, which plays no particular role as long as
the system is strongly degenerate.

One might be tempted at this point to conclude that the generalized Beth
Uhlenbeck formula is valid for arbitrary degeneracy of a fermionic system;
nevertheless, in the next section, we will see that the factor $a_{U}$
itself may become very large at very low temperatures, which in turn
increases the expansion parameter and, automatically, limits the validity of
the expansion at some point.\ We will assume that this somewhat unexpected
phenomenon is a precursor of BCS\ pair condensation; if this is the case,
the low temperature limit of the validity of the generalized Beth Uhlenbeck
is that the gas should remain a normal Fermi gas and, even, not to be too
close to condensation. In general, for bosons as well as for fermions, we
can therefore conclude that our calculations remain valid as long as Bose
Einstein or BCS condensation is not too close.

.

\section{Matrix elements of $\overline{U}_2$\label{elements}}

We now study in more detail the values of the basic ingredient that we use
to describe the effects of the interactions on the thermodynamic properties
of a quantum gas, namely the diagonal matrix element of $\overline{U}%
_{2}^{S,A}$ -- together with those of the un-symmetrized operator if the
particles have spins, see section \ref{spin1/2}. A natural question then is
the following: to determine these coefficients, is it sufficient to
characterize the potential in terms of its binary collision phase shifts,
which determine the asymptotic behavior of interacting wave functions, or is
it also necessary to include some information on the behavior of the wave
functions at short relative distances, inside the potential?\ For bosons at
very low temperatures, is it possible, even more simply, to reason in terms
of the scattering length only?

We have already mentioned in the introduction that, in the literature on low
temperature dilute systems, it is often considered as physically obvious
that the short range properties of the potential play no role in the
thermodynamics; it is then possible to replace the real potential by an
effective potential (or pseudopotential) which has the expression already
given in (\ref{veff}) in terms of the scattering length $a$. With a
potential having zero range, it becomes of course meaningless to take into
account the distortions of the wave functions inside the potential, in other
words the contribution of particles ``in the middle of a collision''.\ But
with real potentials, it is not obvious that these contributions are non
existent, in particular since statistical effects arising from particle
exchange are expected to be more important when the particles are close.\ We
discuss here this question in the context of the generalized Beth Uhlenbeck
formula that we have obtained, with the help of simple examples such as a
step like potential mimicking a real interaction potential, or even hard
spheres.\ We first discuss the Ursell length $a_{U}(k)$, which we have
already introduced above, and which depends on the temperature while $a$
does not.

\subsection{Ursell length\label{Ur-length}}

If the interaction potential $V_{int}$ were sufficiently weak, we could use
a first order perturbation expression of the Ursell operator: 
\begin{equation}
U_2=-\int_0^\beta d\beta ^{^{\prime }}\text{e}^{-\beta ^{^{\prime
}}H_0}V_{int}e^{(\beta ^{^{\prime }}-\beta )H_o}+...  \label{u2premier}
\end{equation}
which would lead to the following expression: 
\begin{equation}
<{\bf k}\mid \overline{U}_2\mid {\bf k}>=-\beta <{\bf k}\mid V_{int}\mid 
{\bf k}>+...  \label{premier}
\end{equation}
Inserting (\ref{veff}) into this result provides: 
\begin{equation}
<{\bf k}\mid \overline{U}_2\mid {\bf k}>=-\beta \frac{4\pi \hbar ^2a}{m%
%TCIMACRO{\TeXButton{V}{\vol} }
%BeginExpansion
\vol%
%EndExpansion
}=-\frac{2a\lambda _T^2}{%
%TCIMACRO{\TeXButton{V}{\vol} }
%BeginExpansion
\vol%
%EndExpansion
}  \label{exp1}
\end{equation}
where $%
%TCIMACRO{\TeXButton{V}{\vol}}
%BeginExpansion
\vol%
%EndExpansion
$ is the macroscopic volume and $\lambda _T$ the thermal wavelength (\ref
{lambda}).

By analogy with this first order calculation, it is convenient to
characterize the diagonal matrix elements in terms of a length, $a_{U}(k)$,
which we call the Ursell length and which we have defined in (\ref{sigma})
as: 
\begin{equation}
a_{U}(k)=-\frac{%
%TCIMACRO{\TeXButton{V}{\vol} }
%BeginExpansion
\vol%
%EndExpansion
}{2\lambda _{T}^{2}}<{\bf k}\mid \overline{U}_{2}\mid {\bf k}>
\label{longursell}
\end{equation}
This very definition ensures that, within the theory of the pseudopotential, 
$a_{U}(k)$ is exactly equal to the scattering length $a$. This length
therefore provides a convenient tool for a discussion of the validity of
this approximation: as long as $a_{U}(k)$ remains very close to $a$ for all
relevant values of ${\bf k}$, the theory of pseudopotentials and ours
provide strictly equivalent results. For identical particles, what is needed
is the symmetrized Ursell length, already defined in (\ref{sigmaprime}) and (%
\ref{sigmaprim}), which allows us to write the correction to the grand
potential for spinless particles in the form (\ref{eq-base}).\ This equation
(\ref{eq-base}) expresses that the symmetrized Ursell length gives, within
the numerical factor in front of this integral, the crossed contribution of
two velocity classes ${\bf k}_{1}$ and ${\bf k}_{2}$ to the grand potential
of the system (its pressure). If the particles have spins, we have to use (%
\ref{Z-spins}) and (\ref{sigma-spins}).

In section \ref{rotation}, we introduced spherical orbital variables from
the free spherical waves $\mid j_{k,l,m}^{(0)}>$.\ At this stage, it is
convenient to use spherical waves that are normalized in a sphere of volume $%
%TCIMACRO{\TeXButton{V}{\vol}}
%BeginExpansion
\vol%
%EndExpansion
=4\pi R^{3}/3$.\ We therefore introduce a new notation, $\mid \varphi
_{klm}^{(0)}>$, but these kets are simply proportional to the $\mid
j_{klm}^{(0)}>$: 
\begin{equation}
\mid \varphi _{klm}^{(0)}>=x_{kl}\mid j_{klm}^{(0)}>  \label{xklm}
\end{equation}
where $x_{kl}$ is a normalization coefficient which in the limit of large
values of the product $kR$ is equal to: 
\begin{equation}
x_{kl}=\sqrt{\frac{2k^{2}}{R}}  \label{tata}
\end{equation}
Introducing in the same way the normalized kets $\mid \varphi _{k^{^{\prime
}}lm}>$ in the presence of the interaction potential, we can write: 
\begin{equation}
a_{U}^{(l)}(k)=-\frac{2\pi }{\lambda _{T}^{2}}\left[ x_{kl}\right]
^{-2}\left[ \sum_{k^{^{\prime }}}\left| <\varphi _{k,l,m}^{(0)}\mid \varphi
_{k^{^{\prime }},l,m}>\right| ^{2}e^{-\beta \left[ e(k^{^{\prime
}})-e(k)\right] }-1\right]  \label{toto}
\end{equation}
where $e(k)$ and $e(k^{^{\prime }})$ are the energies of the free and
interacting states.

In what follows we discuss the values of the Ursell length and its ${\bf k}$
dependence with the help of a few examples.

\subsection{Steplike potential}

In order to simplify our discussion, we now consider an interaction
potential made of a hard core of diameter $xb$ (with $x\leq 1$) and of an
attractive part ($V_{int}=-V_{0}$) from relative distance $r=xb$ to $r=b$;
see figure 1.\ Our discussion is in fact more general, but this kind of
simplified potential is a convenient way to mimic the effects of a more
realistic interaction potential, containing strong repulsion at short
distances and Van der Waals attraction at large distances. For instance, it
is not difficult to find a relation between $V_{0}$, $x$ and $b$ which
ensures that the scattering length of this potential vanishes (compensation
of the effects of attraction and repulsion at low energies). Does this imply
that the matrix elements of $U_{2}$, that is the Ursell length $a_{U}(k)$,
also vanish?\ Not in general, since the scalar products $<\varphi
_{k,l,m}^{(0)}\mid \varphi _{k^{^{\prime }},l,m}>$ in (\ref{toto}) are not
only sensitive to the changes of the interacting wave functions outside of
the potential (which do not occur if the scattering length vanishes) but
also to their values inside the potential.\ In other words, for a degenerate
dilute gas, all the effects of the potential are not necessarily contained
in the scattering phase shifts.

We nevertheless note that, when the range $b$ of the potential tends to
zero, the corrections to the scalar products are necessarily of third order
in $b$ (at least), while first order terms occur in the phase shifts (for
instance, at very low temperatures, the Beth Uhlenbeck formula predicts a
correction which is proportional to $b\lambda _{T}^{2}$). Therefore, if the
potential range is sufficiently small (compared to the two other microscopic
distances, the average distance between particles $n^{-1/3}$ and $\lambda
_{T}$), one may limit the calculation to first order in $b$ so that the
contribution of the distortion of the wave functions inside the potential
may be ignored.\ Within this approximation, all effects of the potential on
the thermodynamic properties are indeed contained in the collision phase
shifts. A brief similar discussion in terms of ``shape dependent terms'' (as
opposed to phase shift dependent terms) appears in the two last paragraphs
of section 5 of ref. \cite{Huang-Yang-Lutt}

Consequently, in all calculations of the thermodynamic quantities which are
limited to first order in the potential range, it is sufficient to
characterize the potential by its long distance effects on the wave function
(phase shifts) only, while short range effects are irrelevant; if all
collisions take place at very low energies, all effects are then contained
in the scattering length only. On the other hand, this is not necessarily
the case if higher order terms in the potential range are included, and
corrections which originate ``inside the potential'', corresponding to the
contribution of particles ``in the middle of a collision'', may appear; see
also \cite{Huang-Yang-Lutt}. Actually, the question remains open as to
whether is would be possible, by some mathematical transformation, to
express these ``in potential effects'' in terms of the phase shifts only;
this is for instance possible in the absence of the phase occupation
factors, since one then gets the usual Beth Uhlenbeck formula for which such
a transformation is known.\ In the presence of these factors, we have made
efforts to investigate the possibility that a similar simplification takes
place, but we have not been able to prove it. A possible conjecture is that
short range and long range effects are in general independent from each
other, but this remains to be proved by a precise example.

\subsection{Hard spheres}

For hard spheres, the interacting wave function does not penetrate into the
potential; this is a special case where all physical effects of the
potential are necessarily contained in the collision phase shifts.\ The
potential range $b$ coincides in this case with the scattering length $a$.\
But this is not sufficient to ensure that the theory of pseudopotentials
should be equivalent to our results, and we now discuss this question.

\subsubsection{Analytical calculation}

In order to obtain the correction to the thermodynamic potential for hard
spheres, we now perform a calculation of the scalar products which appear in
(\ref{toto}).

\paragraph{s waves}

For free waves, the quantification for the wave numbers is given by: 
\begin{equation}
k=\frac{n\pi }R  \label{en}
\end{equation}
while, for hard spheres of diameter $a$, the interacting waves satisfy the
relation: 
\begin{equation}
k^{^{\prime }}=\frac{n^{^{\prime }}\pi }{R-a}  \label{enprime}
\end{equation}
where $n$ and $n^{^{\prime }}$ are integer numbers.\ The scalar products of (%
\ref{toto}) then become functions of these numbers: 
\begin{equation}
<\varphi _{k,0,0}^{(0)}\mid \varphi _{k^{^{\prime }},0,0}>=\frac 2{\sqrt{%
R\left( R-a\right) }}\int_a^Rdr\sin \left( \frac{n\pi r}R\right) \sin \left( 
\frac{n^{^{\prime }}\pi \left( r-a\right) }{R-a}\right)  \label{pscal}
\end{equation}
which, after a simple calculation, leads to: 
\begin{equation}
\left| <\varphi _{k,0,0}^{(0)}\mid \varphi _{k^{^{\prime }},0,0}>\right|
^2=\pi ^{-2}\sin ^2\frac{n\pi a}R\frac{4n^{^{\prime }2}\left[ 1-\frac
aR\right] }{\left[ n^{^{\prime }}-n\left( 1-a/R\right) \right] ^2\left[
n^{^{\prime }}+n\left( 1-a/R\right) \right] ^2}  \label{pscalbis}
\end{equation}

This result, when inserted into (\ref{toto}) for $l=0$, provides the
contributions of s waves to the Ursell length. Equivalent results for
infinite volume (only s wave) are given by Lee and Yang \cite{Lee-Yang1}; we
have checked that our numerical results are compatible with those of these
authors. Since $na/R=ka/\pi $, equation (\ref{pscalbis}) shows that the
scalar product is peaked around a value of $n^{^{\prime }}$ given by: 
\begin{equation}
n^{^{\prime }}=n-E(\frac{ka}\pi )  \label{entiere}
\end{equation}
where $E$ is the integer value; for low energies, $ka\ll 1$ and the
preceding equation reduces to $n^{^{\prime }}=n$.

Assume for instance that we are interested in calculating the Ursell length
to first order in $a$, and in the s wave channel only.\ Equation (\ref
{pscalbis}) shows that only the term $n^{^{\prime }}=n$ contributes and that
the corresponding value is: 
\begin{equation}
\displaystyle\sin ^{2}\frac{n\pi a}{R}\frac{4n^{2}\left[ 1-\frac{a}{R}%
\right] }{\left[ \frac{n\pi a}{R}\right] ^{2}n^{2}\left[ 2-\frac{a}{R}%
\right] ^{2}}\simeq 1  \label{un}
\end{equation}
so that (\ref{toto}) becomes to this order: 
\begin{equation}
a_{U}^{(0)}(k)=-\frac{2\pi }{\lambda _{T}^{2}}\frac{R}{2k^{2}}\left[ -\beta 
\frac{n^{2}\hbar ^{2}\pi ^{2}}{m}\left( \frac{1}{\left( R-a\right) ^{2}}-%
\frac{1}{R^{2}}\right) \right] =a  \label{a}
\end{equation}
The first order value is therefore merely equal to $a$; but we will see that 
$a_{U}^{(0)}(k)$ may strongly differ from this value if higher orders
effects in the potential range are included.

\paragraph{larger $l$ values}

Formula (\ref{pscal}) can be generalized to non-zero angular momentum $l$.
The sine functions of the free and of the hard-sphere wave-functions have to
be replaced by a spherical Bessel function and a linear combination of this
function and a Neumann function of order $l$ respectively. After
normalization we obtain for $|{\bf r}|>a$: 
\begin{equation}
\varphi _{k^{^{\prime }},l,0}({\bf r})=\sqrt{\frac{2a^{2}k^{4}}{R\left[
\,(ka)^{2}m_{l}^{2}(ka)-a/R\right] }}\left[ j_{l}(k^{^{\prime
}}a)n_{l}(k^{^{\prime }}r)-n_{l}(k^{^{\prime }}a)j_{l}(k^{^{\prime
}}r)\right] \,Y_{l}^{0}(\hat{r})  \label{hseigenf}
\end{equation}
$m_{l}(k^{^{\prime }}a)$ stands for ``modulus''\cite[9.2.17]{Abramowitz}: 
\begin{equation}
m_{l}(k^{^{\prime }}a)=\sqrt{j_{l}^{2}(k^{^{\prime
}}a)+n_{l}^{2}(k^{^{\prime }}a)}  \label{hsml}
\end{equation}

The quantification conditions have to be modified; (\ref{enprime}) now
becomes implicit: 
\begin{equation}
k^{^{\prime }}R+\eta _l(k^{^{\prime }})=n\pi +\frac{l\pi }2
\label{hsquantif}
\end{equation}
where $\eta _l(k^{^{\prime }})$ stands for the $l$-wave phase shift for wave
number $k^{^{\prime }}$, which is given by: 
\begin{equation}
\eta _l(k^{^{\prime }})=\arctan \frac{j_l(k^{^{\prime }}a)}{n_l(k^{^{\prime
}}a)}-m_{k^{^{\prime }}}\pi  \label{hsphshift}
\end{equation}
In this definition we limit the values of the inverse tangent to the
interval $\left[ -\frac \pi 2;\frac \pi 2\right[ $ and $m_{k^{^{\prime }}}$
counts the number of times that --- by increasing $k^{^{\prime }}a$ from
zero to the final value --- the value of inverse tangent jumps from $\frac
\pi 2$ to $-\frac \pi 2$. In this way we obtain $\eta _l$ as a continuous
function of $k^{^{\prime }}$. For large values of $k^{^{\prime }}a$, the
effects of the centrifugal barrier become negligible and the quantification
condition becomes independent of $l$ so that it reduces to the $s$-wave
expression (\ref{enprime}).

The normalized free wave function is given by (\ref{besel}), so that we
arrive at the following expression for the square of the scalar product: 
\begin{equation}
\left| \left\langle \varphi _{k\,\,l\,0}^{(0)}|\varphi _{k^{\prime
}\,\,l\,0}\right\rangle \right| ^2=\left[ \frac{2a(ka)(k^{\prime }a)}{%
R\left[ (k^{\prime }a)^2-(ka)^2\right] }\right] ^2\,\frac 1{(k^{\prime
}a)^2m_l^2(k^{\prime }a)-a/R}\,\,j_l^2(ka)  \label{hsscprod}
\end{equation}

\subsubsection{Numerical results\label{numerical}}

Inserting (\ref{pscalbis}) and (\ref{hsscprod}) into (\ref{toto}) and then
into (\ref{sigma}) - or (\ref{sigmaprim}) - provides the Ursell length,
which in turn determines the diagonal matrix element of $\overline{U}_{2}$ -
or $\overline{U}_{2}^{S,A}$ for identical particles.\ For brevity, we just
give the results of our numerical calculations of $a_{U}(k)$; in other words
we only discuss interactions between particles in different spin states if
the particles are indistinguishable. But there is no difficulty in treating
the general case, since the calculation of $a_{U}^{S,A}(k)$ is very similar;
moreover, as soon as the wave number $k$ is sufficiently small, one simply
has $a_{U}^{S}(k)\simeq 2a_{U}(k)$ and $a_{U}^{A}(k)\simeq 0$.\ The results
concerning $a_{U}(k)$ are shown in figures 2 and 3 (more details about the
calculations can be found in \cite{These Peter}).\ There are two parameters
in the problem, $a$ and $\lambda _{T}$ . As convenient dimensionless
variables we choose, either the product $k\lambda _{T}$, or $ka$ ; the ratio 
$a/\lambda _{T}$ is then kept as a fixed parameter, which has small values
either when the potential range is very small or when the temperature is
low.\ Figure 2 shows that, when $a\ll \lambda _{T}$, and as long as $%
k\lambda _{T}$ remains smaller than 1, the Ursell length is indeed equal to $%
a$ (with a good accuracy), so that the theory of the pseudopotential is
fully justified. For higher temperatures, we note that the low energy values
of $a_{U}(k)$ become noticeably different from $a$. This is not surprising
since, when $T$ increases, we progressively reach a classical regime of
small wave packets; they no longer have a much larger spatial extent than
the potential, while this is necessary for the approximation of the
pseudopotential to apply. We therefore concentrate on low values of the
ratio $a/\lambda _{T}$.

If the system is made of bosons, since the distributions $f$'s contained in (%
\ref{eq-base}) have at most the same width as Gaussian thermal exponentials,
the values of $k\lambda _{T}$ that are relevant in the integral are
comparable to 1, or smaller; figure 2 then shows that the use of our theory
or of the theory of pseudopotentials leads to the same results; see the
discussion of section \ref{rotation}, in particular formula (\ref{variation}%
). But assume now that the system is made of fermions\footnote{%
We are dealing here with the description of interactions of fermions in
opposite spin states.}, and that the temperature is progressively lowered at
constant density.\ If the system is degenerate, the width of the functions $%
f $'s is determined by the Fermi momentum $k_{F}$, which in turn depends on
the density; in other words the width is practically independent of the
temperature.\ Meanwhile, if the temperature is lowered more and more, $%
\lambda _{T}$ increases so that the product $k\lambda _{T}$ can take
arbitrarily large values inside the integral.\ On the other hand, $ka$
remains limited to values smaller than 1 since, for degenerate gases where
the distance between the particles is comparable to the inverse Fermi
momentum in a degenerate gas, the diluteness condition $n^{1/3}a\ll 1$
yields: 
\begin{equation}
1\gg k_{F}a\geq ka  \label{ka}
\end{equation}
The departure from 1 of the curves of figure 2, which fall well below this
value when $k\lambda _{T}$ increases, shows that significant discrepancies
from a pseudopotential theory may indeed be obtained. This is even more
visible in figure 3, which uses a different variable, the product $ka$: even
for small values of the ratio $a/\lambda _{T}$, significant departures of
the Ursell length from $a$ are obtained for small values of $ka$; in other
words, (\ref{corrferm}) is no longer a good approximation of the generalized
Beth Uhlenbeck formula. Actually, the smaller $a/\lambda _{T}$, the more
pronounced these departures are; we have an illustration of the consequences
of the temperature dependence of $\overline{U}_{2}$, where the effects of
the potential are more and more delocalized by thermal effects while $%
\lambda _{T}$ increases; consequently, variations of the matrix element
occur even if the range of the potential is very small and even if $ka\ll 1$%
.\ In figure 4, we plot the variations of the diagonal element $<{\bf r}\mid
\left[ U_{2}\right] _{rel}\mid {\bf r}>$ as a function of the relative
position ${\bf r}$; the results show clearly how the second Ursell operator
acquires a longer and longer range at decreasing temperatures.

A striking feature of figure 3 is the change of sign of the Ursell length
which takes place when $ka$ increases from zero. The origin of this change
is understandable from (\ref{toto}), from which one can convince oneself
that the contribution of low values of $k^{^{\prime }}$ becomes dominant as
soon as $k$ is sufficiently large; this is because, while the scalar product 
$\left| <\varphi _{k,l,m}^{(0)}\mid \varphi _{k^{^{\prime }},l,m}>\right| $
with $k^{^{\prime }}\simeq 0$ decreases relatively slowly when $k$ increases
(as $k^{-2}$), the exponential $e^{\beta e(k)-\beta e(k^{^{\prime }})}$
varies much more rapidly and so that it makes small values of $k^{^{\prime
}} $ dominate the sum.\ In other words, what determines the diagonal matrix
element of $U_{2}$ is the contribution of interacting states that have a
very small relative energy; because these states evolve more slowly in time
than the free wave packets of energy $\hbar ^{2}k^{2}/m$, the net
differential result is equivalent to an attraction.\ We therefore come to
the conclusion that even hard cores can result in an effective attraction
between fermions of opposite spins at the surface of a Fermi sphere,
provided that the temperature is sufficiently low.\ But other interesting
features also appear; for instance there is a strong dependence of the
matrix element on the relative momentum and, for some value of $k$, the
Ursell length (the effective interaction) vanishes; probably more important
is the fact that the effective interaction increases almost exponentially as
a function of $k^{2}$, which is nothing but the square of the relative
momentum of the interacting particles.\ This shows that the correction to
the partition function is dominated by processes that take place preferably
between particles having almost opposite momenta on the surface of the Fermi
sphere (assuming that they have opposite spins).

\subsubsection{Validity of the Ursell expansion}

We now come back to the discussion made at the end of section \ref{physdisc}
concerning the validity of the Ursell expansion for a dilute gas of
fermions.\ The result of this discussion was that the expansion parameter is
the product $n^{1/3}a_{U}$. This parameter would remain a constant as a
function of temperature for a gas of constant density if the Ursell length
did not vary too much as a function of temperature, so that the situation
would be simple.\ But in fact, we have actually found for Fermi gases that
the maximum value of $a_{U}$ becomes larger and larger at low temperatures.\
This automatically limits the range of validity of the first order $U_{2}$
theory to temperatures at which the relevant values of the Ursell length are
not too large.

To determine a limit temperature, we will look for an asymptotic expression
of the Ursell length valid for wave vectors of the magnitude of the Fermi
wave vector $k\approx k_{F}$. In this case, the terms corresponding to low
values of $k^{\prime }$ in the sum over the states of interacting pairs of (%
\ref{toto}) are dominant. When $k^{\prime }\ll k_{F}$ the square of the
scalar product (\ref{hsscprod}) reduces to: 
\begin{equation}
\left| \left\langle \varphi _{k_{F}\,\,0\,0}^{(0)}|\varphi _{k^{\prime
}\,0\,0}\right\rangle \right| ^{2}\approx \frac{4}{R^{2}}\left( \frac{%
k^{\prime }}{k_{F}^{2}}\right) ^{2}\sin ^{2}k_{F}a
\end{equation}
We replace the square of the scalar product in (\ref{toto}) by this
expression and transform the sum into an integral: 
\begin{eqnarray}
a_{U}(k_{F}) &\propto &-\frac{R^{2}}{k_{F}^{2}\lambda _{T}^{2}}\frac{1}{%
k_{F}^{4}R^{2}}\sin ^{2}k_{F}a\,\,\text{e}^{\beta e_{k_{F}}}\int \text{d}%
k^{\prime }k^{\prime 2}\text{e}^{-\beta e_{k^{\prime }}} \\
&\propto &-\frac{\lambda _{T}}{(k_{F}\lambda _{T})^{6}}\sin ^{2}k_{F}a\,\,%
\text{e}^{\beta e_{k_{F}}}
\end{eqnarray}
where $e_{F}$ is the Fermi energy (proportional to $n^{2/3}$ for a strongly
degenerate gas). We are interested in the limit $n^{1/3}a\ll 1\,$or $%
k_{F}^{1/3}a\ll 1$; we can thus replace the sine by its first order
approximation, which yields: 
\begin{equation}
\left| a_{U}(k_{F})\right| \propto a\frac{1}{(k_{F}\lambda _{T})^{4}}\frac{a%
}{\lambda _{T}}\text{e}^{e_{F}/k_{B}T}  \label{dependance}
\end{equation}
The validity criterion of the Ursell expansion then becomes: 
\begin{equation}
n^{1/3}a\frac{1}{(k_{F}\lambda _{T})^{4}}\frac{a}{\lambda _{T}}\text{e}%
^{e_{F}/k_{B}T}\ll 1  \label{cond}
\end{equation}
which shows that the theory is valid only when the temperature satisfies the
approximate condition: 
\begin{equation}
k_{B}T_{C}\gtrsim \frac{e_{F}}{\func{Log}\left[ n\lambda _{T}^{3}(\lambda
_{T}/a)^{2}\right] }  \label{validity}
\end{equation}
Inside the logarithmic function in the denominator, both the factor $%
n\lambda _{T}^{3}$ and $(\lambda _{T}/a)^{2}$ are larger than one,
especially the latter for a very dilute gas; but since the logarithm has
slow variations, whatever parameters we choose, in practice the limit
temperature fixed by this condition is not lower than is about a tenth of
Fermi temperature, a rather a high temperature compared to the transition
temperature for Cooper pairing.

Our conclusion is that, for any type of interaction, and even if the density
of the gas is fixed at a very low value, at sufficiently low temperature the
Ursell length becomes larger and larger so that the Ursell expansion is no
longer a good expansion.\ The maximum value of the Ursell length is obtained
for pairs of fermions with opposite spins and maximum relative momenta,
which corresponds to two fermions having opposite momenta on the surface of
the Fermi sphere; this is reminiscent of the BCS pairing phenomenon, while
the phenomenon takes place at much higher temperatures (the right hand side
of (\ref{validity}) does not coincide with the standard expression of the
critical temperature) so that stricto sensu it can not be called a precursor
of this transition. Whether of not it is related to this transition, the
change of sign of the Ursell length will have a strong effect on the two
body correlation function in the system (we have seen in \cite{II} that the
second Ursell operator is closely related to the short range properties of
the two body density operator).\ These results are also reminiscent of the
well known work of Luttinger and Kohn \cite{LK} who predict the occurrence
of superconductivity in purely repulsive systems, as well of the more recent
work of Kagan and Chubukov \cite{KC} who also predict in this case $p$ wave
superfluidity in a dilute Fermi gas. We are planning to investigate this
connection in more detail in a future article.

To summarize, for both fermionic of bosonic systems, the validity of the
Ursell expansion is limited, even for a very dilute gas, to temperatures
sufficiently above any transition point.\ Mathematically, in the case of
bosons the divergence of the series arises from the distribution function
(or operator) $f$, while in the case of fermions its origin is the
increasing value of the matrix elements of $U_{2}$ themselves.

\section{Conclusion}

Our formalism provides a systematic treatment of the interactions in a
dilute gas where the basic objects are not the matrix elements of the
potential itself but those of temperature dependent operators. In \cite{II},
we have shown how microscopic, short range correlations between particles
could explicitly be taken into account and calculated. In the present
article, we investigate the macroscopic properties of the gas by basing the
discussion on expression (\ref{logz-int}), which resembles a first order
perturbative expression of an energy correction, while it actually is rather
different.\ This is mostly because the matrix element which appears in the
expression is not the matrix element of the potential itself, or of some
variety of pseudopotential, but the matrix element of the second Ursell
operator $U_{2}$, which corresponds physically to a local Boltzmannian
equilibrium.\ Except of course in trivial cases where the interaction
potential is indeed weak for all values of the relative distances of the
particles, which allows for a first order treatment of the potential, this
introduces significant differences; the major reason is the temperature
dependence of the matrix elements of $U_{2}$, which is in general more
complex than being merely proportional to $\beta $ (as would be the case in
usual perturbation theory). This matrix element is conveniently described in
terms of the Ursell length.\ We have seen in particular that, for fermions
at very low temperatures, effective attractions at the surface of a Fermi
sphere may take place, independently of the repulsive or attractive
character of the potential itself.\ Moreover, our formalism contains
naturally effects such as the statistical exchange between bound molecules
and free particles, which may play some role in experiments with alkali
atoms at very low temperatures \cite{Cornell-1995,Bradley-1995,Davis-1995}.

\newpage

\begin{center}
Acknowledgments
\end{center}

Many useful and friendly discussions with Roger Balian took place on the
subject of this article.\ A.M. and W.M.\ would like to thank the Kastler
Brossel group for hospitality during stays in Paris.\ We are grateful for
the following financial supports: NSF\ grant DMR-9412769 (A.M.) and
INT-9015836 in conjunction with a CNRS grant (F.L. and A.M.), and NATO%
\TEXTsymbol{\backslash}OTAN CRG 930644 (F.L.\ and W.M.).

\begin{center}
Note added
\end{center}

After completion of this manuscript, we received a preprint from H.\ Stein
et al.\cite{Stein} who, in the case of bosons, find that the many body
scattering length undergoes a divergence when the superfluid transition is
approached, an effect similar to what we obtain for fermions.\ The
significance of this similarity has not been examined in detail yet.

\begin{center}
Figure captions
\end{center}

Figure 1: Steplike potential used to discuss compensation effects between
the attractive and the repulsive part on the scattering length; $V_{0}$ is
the depth of the attractive potential, $b$ its range; a hard core of range $%
xb$ is assumed ($x<1$). While a relation between these three parameters can
be found to make the scattering length $a$ vanish, the effect of the
potential on the wave function inside its range remains significant.

Figure 2: Variations of the ratio between the Ursell length $a_U(k)$ and the
range $a$ of the hard core potential as a function of the dimensionless
variable $k\lambda _T$.\ The ratio $a/\lambda _T$, with values shown in the
upper right, is a parameter which takes small values if either the
temperature is sufficiently low or the radius of the hard cores sufficiently
small.

Figure 3: Variations of the same quantity as in Figure 2, but as a function
of the dimensionless variable $ka$; this representation is convenient for
the discussion of fermions (where it characterizes interactions between
particles in opposite spin states).

Figure 4: Variations of the diagonal matrix element $<{\bf r}\mid \left[
U_{2}\right] _{rel}\mid {\bf r}>$ as a function of ${\bf r}$, for various
values of the parameter $a/\lambda _{T}$; at low temperatures, the thermal
increase of the range of the operator is clearly visible.

\newpage\

\end{document}